\documentclass[twocolumn,superscriptaddress]{revtex4}%%,twocolumn
\usepackage{graphicx}
\usepackage{epstopdf}
\usepackage{times}
\usepackage{amsmath}
\usepackage{amssymb}
\usepackage{units}
\usepackage{stmaryrd} %% "// sign" = \sslash
\usepackage[compact]{titlesec}
\titlespacing{\section}{0pt}{*0}{*0}
\titlespacing{\subsection}{0pt}{*0}{*0}
\titlespacing{\subsubsection}{0pt}{*0}{*0}

\makeatletter
\newcommand*{\rom}[1]{\expandafter\@slowromancap\romannumeral #1@}
\makeatother

\begin{document}
\preprint{0}

\title{Comparative study of rare earth hexaborides using high resolution angle-resolved photoemission}

\author{S. V. Ramankutty}
\email{s.v.ramankutty@uva.nl} 
 \address{Van der Waals-Zeeman Institute, Institute of Physics (IoP), University of Amsterdam, Science Park 904, 1098 XH, Amsterdam, The Netherlands}

\author{N. de Jong}
\address{Van der Waals-Zeeman Institute, Institute of Physics (IoP), University of Amsterdam, Science Park 904, 1098 XH, Amsterdam, The Netherlands}

\author{Y. K. Huang}
\address{Van der Waals-Zeeman Institute, Institute of Physics (IoP), University of Amsterdam, Science Park 904, 1098 XH, Amsterdam, The Netherlands}

\author{B. Zwartsenberg}
\thanks{Current adress: Department of Physics and Astronomy, University of British Columbia, Vancouver, British Columbia V6T 1Z1, Canada}
\address{Van der Waals-Zeeman Institute, Institute of Physics (IoP), University of Amsterdam, Science Park 904, 1098 XH, Amsterdam, The Netherlands}

\author{F. Massee}
\thanks{Current address: Laboratoire de Physique des Solides, Universit\'{e} Paris-Sud}
\address{Laboratory of Atomic and Solid State Physics, Department of Physics, Cornell University, Ithaca, NY 14853, United States of America}

\author{T. V. Bay}
\thanks{Current address: Zernike Institute for Advanced Materials, University of Groningen}
\address{Van der Waals-Zeeman Institute, Institute of Physics (IoP), University of Amsterdam, Science Park 904, 1098 XH, Amsterdam, The Netherlands}

\author{M. S. Golden}
\email{m.s.golden@uva.nl} 
\address{Van der Waals-Zeeman Institute, Institute of Physics (IoP), University of Amsterdam, Science Park 904, 1098 XH, Amsterdam, The Netherlands}

\author{E. Frantzeskakis}
\email{e.frantzeskakis@uva.nl} 
 \address{Van der Waals-Zeeman Institute, Institute of Physics (IoP), University of Amsterdam, Science Park 904, 1098 XH, Amsterdam, The Netherlands}
 
\date{\today}

\begin{abstract}
Strong electron correlations in rare earth hexaborides can give rise to a variety of interesting phenomena like ferromagnetism, Kondo hybridization, mixed valence, superconductivity and possibly topological
characteristics. The theoretical prediction of topological properties in SmB$_{6}$ and YbB$_{6}$, has rekindled the scientific interest in the rare earth hexaborides, and high-resolution ARPES has been playing a major role in the debate. The electronic band structure of the hexaborides contains the key to understand the origin of the different phenomena observed, and much can be learned by comparing the experimental data from different rare earth hexaborides. We have performed high-resolution ARPES on the (001) surfaces of YbB$_{6}$, CeB$_{6}$ and SmB$_{6}$. On the most basic level, the data show that the differences in the valence of the rare earth element are reflected in the experimental electronic band structure primarily as a rigid shift of the energy position of the metal 5$\textit{d}$ states with respect to the Fermi level. Although the overall shape of the $\textit{d}$-derived Fermi surface contours remains the same, we report differences in the dimensionality of these states between the compounds studied. Moreover, the spectroscopic fingerprint of the 4$\textit{f}$ states also reveals considerable differences that are related to their coherence and the strength of the $\textit{d}$-$\textit{f}$ hybridization. For the SmB$_6$ case, we use ARPES in combination with STM imaging and electron diffraction to reveal time dependent changes in the structural symmetry of the highly debated SmB$_{6}$(001) surface.
All in all, our study highlights the suitability of electron spectroscopies like high-resolution ARPES to provide links between electronic structure and function in complex and correlated materials such as the rare earth hexaborides. \end{abstract}

\maketitle

\section{INTRODUCTION}

Over the past decades, rare earth hexaborides -i.e. REB$_{6}$, where RE is a rare earth element- have attracted considerable interest of the scientific community because electron correlations can give rise to unconventional properties and exotic ground states \cite{Aeppli1992,Riseborough2000,Coleman2007}.
Typical examples are the debate on excitonic ferromagnetism in divalent hexaborides doped with RE elements \cite{Young1999,Barzykin2000,Zhitomirsky1999}, the dense Kondo behavior of CeB$_{6}$ \cite{KomatsubaraJMMM1983,NakamuraJPSJ1995} and mixed valent SmB$_{6}$ as an archetypal heavy fermion, Kondo insulator system \cite{SoumaPhysicaB2002,Nozawa2002} and as a candidate for the first correlated topological insulator \cite{Dzero2010,FengLu2013}. Variations in the occupancy, and hence the energy position, of both the RE \textit{d} bands and the localized \textit{f} states among different compounds are responsible for the plethora of properties these systems display. Focusing on these variations, this paper presents a comparative study of three RE hexaborides that have been heavily re-investigated during the last 5 years by means of theoretical calculations and modern spectroscopic techniques: divalent YbB$_{6}$, trivalent CeB$_{6}$ and mixed valent SmB$_{6}$.

The theoretical prediction that SmB$_{6}$ could be a topological Kondo insulator led to renewed and intense scientific interest in RE hexaborides. The Kondo hybridization gap that arises from the interaction of localized \textit{f} electrons and delocalized conduction carriers (\textit{d} electrons) in SmB$_{6}$ was predicted to host surface states of non-trivial character \cite{Dzero2010,FengLu2013,Alexandrov2013}. This prediction gave rise to numerous transport \cite{Wolgast2013,Li2013,Fisk2013,Zhang2013,KimNMat}, scanning tunneling microscopy/spectroscopy (STM/STS) \cite{Hoffmanarxiv2013,Rossler2014,Ruan2014} and angle-resolved photoelectron spectroscopy (ARPES) studies \cite{HasanSmB6,FengSmB6,EmmanouilSmB6,DamascelliSmB6,Denlingerarxiv2013,Reinert2014,EmileSmB6} that sought for signatures of topological characteristics in SmB$_{6}$.
At low temperatures (under 5K), various transport experiments \cite{Wolgast2013,Fisk2013,Zhang2013,KimNMat,Syers2015} and torque magnetometry \cite{Li2013} data point to the co-existence of a robust surface-related conduction channel in parallel to the bulk channel that dominates at higher temperatures.
Additional characteristics such as the sensitivity to magnetic impurities \cite{KimNMat} and the Berry phase from the Landau levels seen in quantum oscillations \cite{Li2013} also point towards a topologically non-trivial origin for the surface conduction in SmB$_6$.    
In the Bi-based topological insulators, there is unanimity in the field that ARPES experiments (conducted by myriad groups on as many crystals of multiple material systems) have shown the existence of the tell-tale Dirac cone dispersion of the 2D topological surface states (TSS). The same unanimity goes for STS data from Bi-based topological insulators, in which Landau level spectroscopy at the surface and quasiparticle interference experiments signal the topologically non-trivial nature of the surface states.
The situation is different for SmB$_6$. Here there is no unanimity: the ARPES results are interpreted differently in different publications and it is evident that the (001) cleavage surface of SmB$_6$ has a complicated structure  \cite{Hoffmanarxiv2013,Rossler2014,Ruan2014}. This means that although the (magneto)transport data may represent a smoking gun for topological Kondo insulator behavior in this system, the debate on the true meaning of data from surface-sensitive probes is still wide open.\\
\indent
Recently, the debate on REB$_{6}$ has been extended to other compounds. Divalent YbB$_{6}$ has also been theoretically predicted to host topological surface states \cite{WengPRL2014}. In this case there are no long-standing puzzles in the transport data that require explanation and the valency of the Yb has long been known to be simply divalent. Some recent ARPES studies support the `topological' claim for YbB$_6$ \cite{MingYbB6,HasanYbB6,FengYbB6}, and one paper contests this conclusion from an ARPES point of view \cite{EmmanouilYbB6}. \\
\indent
Last but not least, very recent ARPES investigations of trivalent CeB$_{6}$ have proposed strong band renormalization and the formation of hotspots on its Fermi surface \cite{Hasanarxiv2014}, while transport studies reported the suppression of the dense Kondo state under pressure \cite{ForoozaniPhysicaB2015}. \\
\indent
These facts all add up to make the experimental investigation of the Sm, Yb and Ce hexaborides something of topical interest. 
The aim of this article in this special issue of the journal is to provide further insights into the electronic properties of SmB$_{6}$, YbB$_{6}$ and CeB$_{6}$ by reviewing the status of the field and by providing new spectroscopic data comparing the different compounds.  
In the following, we kick-off by discussing the RE-dependent evolution of the Fermi surface as measured using ARPES. After discussing the electronic structure related to \textit{d}-derived bands, we will move on to the importance of the energy position, energy-broadening and spectral intensity variation of the localized \textit{f} states. We then go on to provide an experimental viewpoint on the dimensionality of the CeB$_{6}$ Fermi surface, comparing with previous data from SmB$_{6}$ and YbB$_{6}$. Finally, we close by reporting a combined ARPES, STM and low-energy electron diffraction (LEED) investigation on the (001) cleavage surface of SmB$_{6}$ that reveals the complex and time-dependent nature of the crystal termination.\\

\section{EXPERIMENTAL DETAILS}

\subsection{Sample growth and cleavage}

SmB$_{6}$, YbB$_{6}$ and CeB$_{6}$ single crystals were grown in an optical floating zone furnace (Crystal Systems, Inc., FZ-T-12000-S-BUPC) under 5 MPa pressure of high-purity argon gas \cite{Bao2013}. The growth rate was 20 mm/h with the feed and seed rods counter-rotating at 30 rpm. Samples with a (001) surface termination were cleaved at 38 K (ARPES) and 135 K (STM), at pressures lower than 3$\times$10$^{-10}$mbar.\\

\subsection{Scanning tunneling microscopy}

STM measurements were performed using a commercial low temperature instrument from Createc. The sample was transferred rapidly and directly from the pre-cooled sample manipulator (where cleavage was carried out) to the cryogenic UHV of the STM, which is housed in a UHV system with room temperature base pressure below 5.0$\times$10$^{-10}$ mbar. All STM data were acquired at 4K. The thermal drift of the STM across 12h is of the order of only a few {\AA}, spatial ($xy$) resolution is subatomic and in the $z$-direction $<$ 0.05 {\AA}. Pt/Ir tips were used for the measurements after testing them on a Au(788) surface which gave sharp step edges of 2.1 {\AA} height. The topographic images presented in this article were measured with a bias voltage of 200mV and a tunnelling current of 10 pA.\\

\subsection{Angle resolved photoelectron spectroscopy}

ARPES experiments were performed at the UE112-PGM-2a-1$^{2}$ beamline (BESSY II storage ring at the Helmholtz Zentrum Berlin) using a Scienta R8000 hemispherical electron analyzer and a six-axis manipulator. The pressure during measurements was better than 3.0$\times$10$^{-10}$ mbar, and the sample temperature was maintained at 38K. The energy position of the Fermi level was determined for every cleave by evaporation of Au films onto the sample holder, such that the gold film was in direct electrical contact with the crystal. The polarization of the incoming photon beam was linear horizontal and the entrance slits of the hemispherical analyzer were vertical.
The whole surface Brillouin zone was spanned by sequential rotation of the polar angle. Measurements with variable photon excitation energy were used to access the out-of-plane dispersion of the electronic states. The conversion of the photon energy to the relevant $k_{\textmd{z}}$ values was carried out assuming free electron final states and an inner potential V$_{\textmd{0}}$ of 14 eV.\\

\section{RESULTS AND DISCUSSION}

Rare earth hexaborides crystallize in a cubic lattice with RE atoms at the corners of the bulk unit cell and a B$_{6}$ octahedron at its body centre \cite{Fisk1974}. Fig. 1(a) shows a sketch of the crystal structure, the bulk Brillouin zone and its 2D projection on the (001) plane. Before presenting our experimental data on rare earth hexaborides, it is instructive to start with Fig. 1(b), which is a cartoon of the low-energy electronic structure inspired by bulk band structure calculations \cite{Jun2007,Gmitra2014,Langford1988,TrompPRL2001,DenlingerPRL2002}. The purple and green curves denote the dispersing RE 5\textit{d} states and the B 2\textit{p} bands, respectively. The horizontal dashed line is the Fermi energy ($E_{\textmd{F}}$). The solid, horizontal blue line schematically shows the energy position of the uppermost occupied RE \textit{f}-derived multiplet, as derived from experimentally determined values \cite{EmmanouilSmB6,EmmanouilYbB6,Hasanarxiv2014,TrompPRL2001}. We do this, as first principle calculations generally fail to capture the correct energy position of the strongly correlated \textit{f} states. The two labelled panels show the electronic band structure in the bulk of the crystal and on its surface. In the following, we will focus on the electronic band structure around the $\Gamma$ and X high-symmetry points of the bulk Brillouin zone.
It is in these momentum-space (i.e. $k$-space) regions that the dispersive B 2\textit{p} and RE 5\textit{d} bands are predicted to overlap \cite{TrompPRL2001} and clear spectroscopic signatures of dispersing contours have been detected experimentally at the Fermi level \cite{DenlingerPRL2002,DenlingerPhysicaB,EmmanouilSmB6,EmmanouilYbB6,Hasanarxiv2014}.

\begin{figure}
\centering
\includegraphics[width =8 cm]{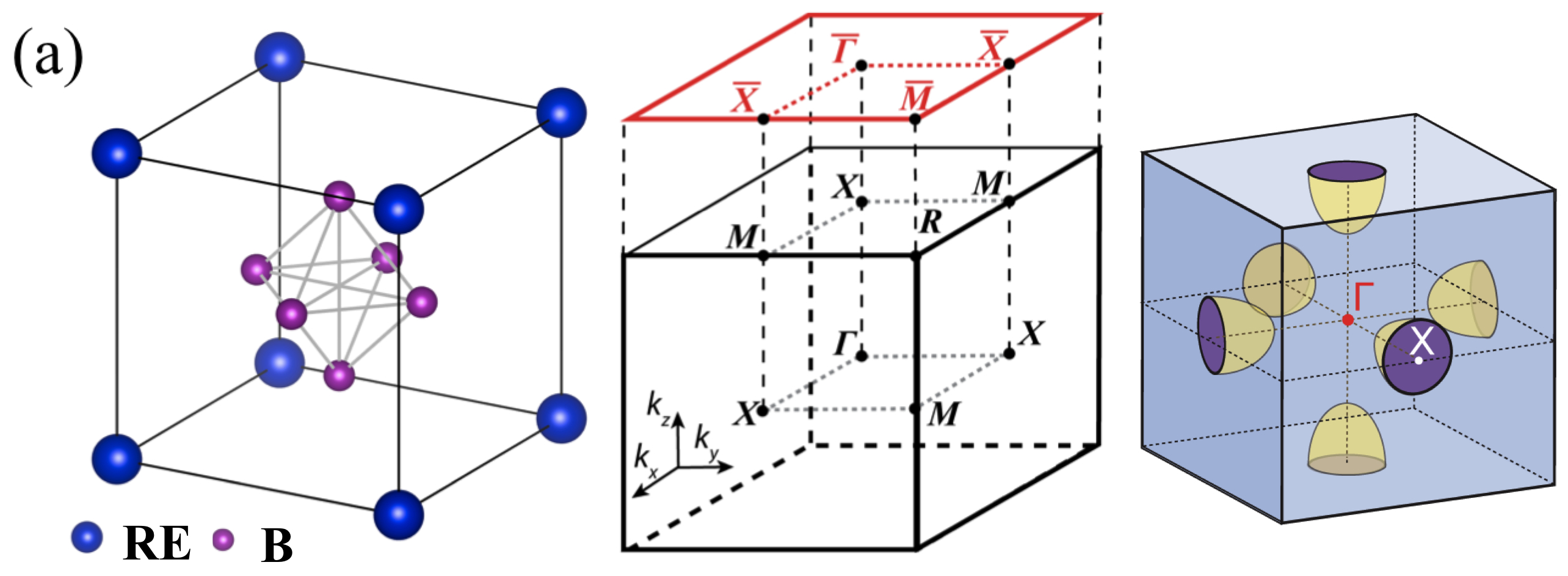}
\includegraphics[width =8 cm]{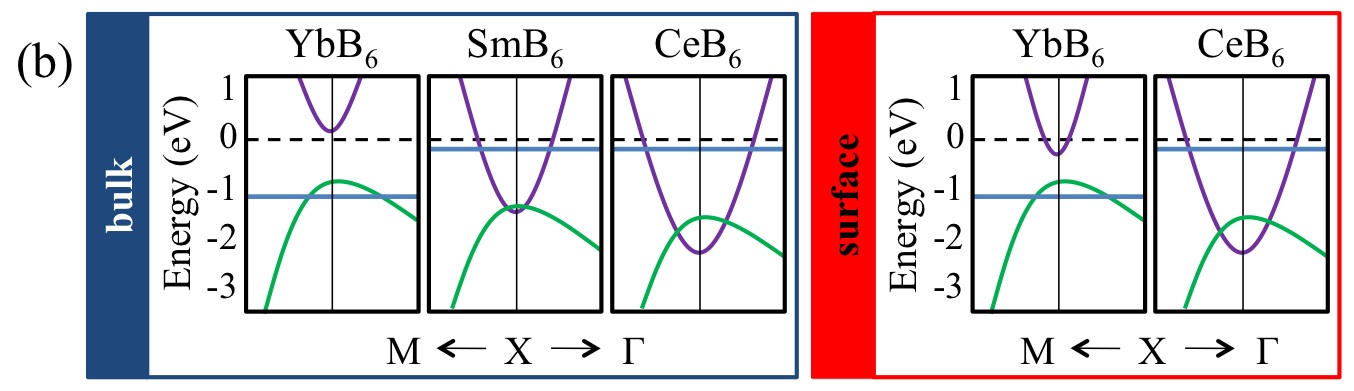}
\includegraphics[width =8 cm]{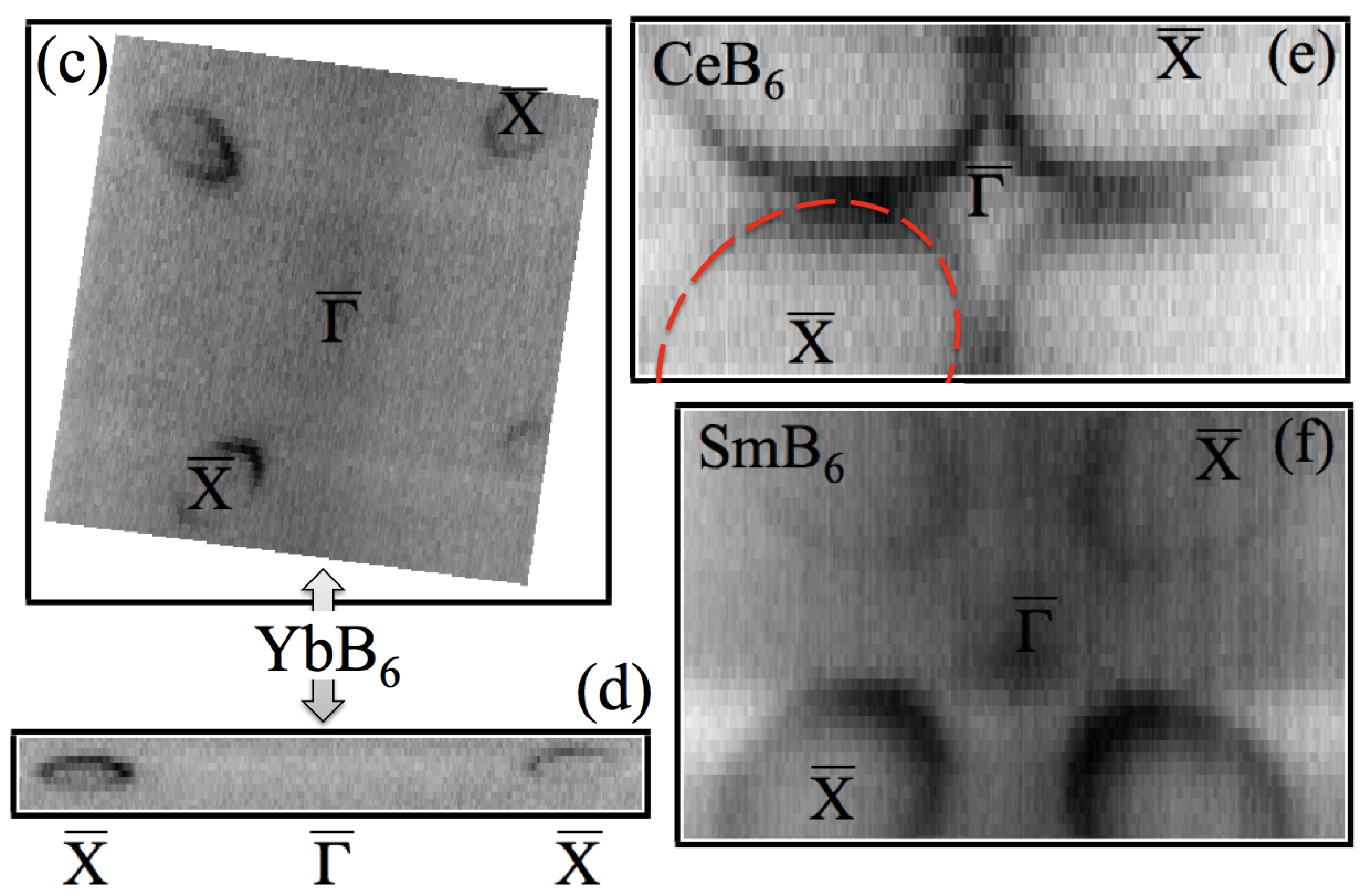}
\caption{(a) Schematic illustrations of (left) the crystal structure of rare earth hexaborides REB$_{6}$; (centre) the bulk Brillouin zone and its projection on the (001) surface; (right) 3D Fermi surface ellipsoids centred at the bulk X-points (adapted from Ref. \onlinecite{Hoffmanarxiv2013}). (b) A simplified schematic version of the main features in the electronic band structure of YbB$_{6}$, SmB$_{6}$ and CeB$_{6}$: blue, purple and green denote the uppermost occupied rare earth \textit{f}-derived multiplet, the rare earth 5\textit{d}-derived band and the B 2\textit{p}-derived band, respectively. The horizontal dashed line marks the Fermi energy. The two labelled panels show the electronic band in the bulk and at the surface, respectively. (c) \& (d) Experimental Fermi surfaces of YbB$_{6}$(001) acquired with 35 eV and 70 eV photons respectively. (e) \& (f) Experimental Fermi surfaces of CeB$_{6}$(001) and SmB$_{6}$(001), respectively, both acquired with 70 eV photons. All experimental data were measured at 38K.}
\label{fs}
\end{figure}

Comparing the three band structure sketches on the 'bulk' panel of Fig. 1(b), a first difference is related to the energy position of the uppermost \textit{f}-derived multiplet with respect to the Fermi level. While for SmB$_{6}$ and CeB$_{6}$, an \textit{f} state lies very close to $E_{\textmd{F}}$, the lowest lying occupied \textit{f} state has a binding energy ($E_{\textmd{b}}$) of $ca$. 1 eV in YbB$_{6}$. Such energy variations are a direct consequence of the different occupancy of the \textit{f} states in these rare earth elements: while in Sm and Ce the 4\textit{f} states are only partially occupied (4\textit{f}$^{5}$ and 4\textit{f}$^{6}$ for Sm and 4\textit{f}$^{1}$ for Ce), Yb has a filled 4\textit{f} shell, which pushes the \textit{f} state down to higher binding energy. As a result, the \textit{d}-\textit{f} interaction in the Sm and Ce compounds is more enhanced than in YbB$_{6}$. This gives rise to a dense Kondo system for CeB$_{6}$ \cite{Takase1980} and to a mixed-valent Kondo insulator for SmB$_{6}$ \cite{Khomskii}, although early pressure-dependent transport data showed discrepancies from the expectations of a simple hybridization gap scenario at low temperature \cite{CooleyPRL}.

A second important difference among the three bulk sketches of Fig. 1(b) is the energy position of the Fermi level with respect to the B 2\textit{p} and RE 5\textit{d} bands. We first consider the case of YbB$_{6}$ where the RE element is divalent. 
It has been reported that although YbB$_{6}$ single crystals may show metallic behavior in resistivity measurements at low temperature, defect-free YbB$_{6}$ would behave as an intrinsic semiconductor with a finite energy gap between the 5\textit{d} and 2\textit{p} bands \cite{Tarascon1980}. The metallic behavior of YbB$_{6}$ is then ascribed to the formation of impurity bands \cite{Tarascon1980}.
A similar bulk semiconducting behavior has been observed on ultrapure single crystals of CaB$_{6}$, where Ca is also divalent \cite{Cho,Rhyee}. The absence of \textit{d}-\textit{p} band overlap in divalent hexaborides has been successfully captured by GW calculations \cite{TrompPRL2001,DenlingerPRL2002}. All in all, bulk-sensitive experimental probes \cite{Tarascon1980,DenlingerPRL2002,Cho,Rhyee} and appropriate calculations of the bulk electronic structure \cite{TrompPRL2001,DenlingerPRL2002} suggest that there is a finite energy gap in the bulk electronic structure of YbB$_{6}$ with the bottom of the conduction band just above $E_{\textmd{F}}$. This is what is shown in the left-most bulk electronic structure cartoon shown in
Fig. 1(b). As an intrinsic semiconductor, YbB$_{6}$ is expected to be prone to surface band bending. Consequently, the surface electronic structure will be different than the bulk as has been observed experimentally for other divalent hexaborides \cite{DenlingerPRL2002}. The left-hand side of the `surface' panel of Fig. 1(b) shows a sketch of the surface electronic structure of YbB$_{6}$ as we have directly determined it using ARPES. The surface chemical potential is intersecting
the bulk conduction band and an electron pocket is observed \cite{MingYbB6,HasanYbB6,FengYbB6,EmmanouilYbB6}. We believe that the resulting potential is so strong that accounts for the observed confinement effects
\cite{MingYbB6,HasanYbB6,FengYbB6,EmmanouilYbB6}, exactly as in the case of Bi-based TIs (e.g. Refs. \onlinecite{Bianchi2010,King2011}). Our earlier ARPES study on YbB$_{6}$ \cite{EmmanouilYbB6} has reported a pronounced time dependent
energy shift of these low-lying occupied states towards the unoccupied part of the spectrum, which may be attributed to changes to the initial surface band bending due to the effect of the photon beam and/or changes in adsorption at the surface 
\cite{Kordyuk2011,Frantzeskakis2014,Frantzeskakis2015}.

We now turn our attention to CeB$_{6}$ where the RE element is trivalent. Here, the bulk chemical potential lies at higher energy, crossing the RE 5\textit{d} bands and there is also a
finite \textit{d}-\textit{p} band overlap \cite{Mo2001,Hasanarxiv2014}. Hence, trivalent hexaborides are not semiconducting and no surface band bending is expected to occur. In the case of CeB$_{6}$, the electronic structure in the bulk and at the surface should therefore be identical, as indicated schematically in Fig. 1(b). The absence of band bending and degree of surface confinement of the electronic structure of CeB$_{6}$ will be discussed in detail in the context of Fig. 3. We finish this comparison by noting that the main features of the bulk electronic structure of mixed-valent SmB$_{6}$ fall nicely between divalent (e.g. YbB$_{6}$) and trivalent (e.g. CeB$_{6}$) compounds: the chemical potential lies at an intermediate energy and the B 2\textit{p} bands are barely touching the RE 5\textit{d} bands \cite{Mo2001}.

We reiterate that these band structure sketches are just that, as they only contain the low-energy electronic states, they consider neither band hybridization effects, nor those of electronic correlations, and - like all bulk band structure calculations - they cannot account for electronic states that are localised at or near the surface of the system.
Nevertheless, the message from the relevant calculations \cite{Jun2007,Gmitra2014,Langford1988,TrompPRL2001} is clear: the overall shape of the dispersive RE 5\textit{d} and B 2\textit{p} bands is barely modified from one RE to another, but there {\it is} a continuous shift of the states with respect to the Fermi level as the RE valence changes from divalent for RE=Yb, to the mixed-valence case for RE=Sm and finally to trivalent for CeB$_{6}$. During this evolution, the RE 5\textit{d} band (purple curve), which crosses $E_{\textmd{F}}$ around the X-point in $k$-space, forms an electron pocket which continuously increases in size as the RE valency increases.

Figure1(c)-(f) present ARPES data from the (001) cleavage surfaces of the the three RE hexaborides under consideration. All four data images in panels (c)-(f) show the intensity at E$_F$ (dark = higher intensity), and the labels indicate high symmetry points in the surface Brillouin zone (SBZ) shown in panel (a).
For all compounds, the most prominent Fermi surface features are elliptical electron pockets centered around the $\overline{\textmd{X}}$ points. These contours are the smallest for YbB$_{6}$, shown in panels (c) and (d) and are largest for CeB$_{6}$ [panel (f)]. Similar elliptical contours were observed in early ARPES studies on various metal hexaborides \cite{DenlingerPhysicaB,DenlingerPRL2002} and are attributed to the Fermi surface signature of the 5\textit{d} states. In 3D $k$-space, these Fermi surfaces form ellipsoids centered at all six X high-symmetry points of the bulk Brillouin zone, as shown schematically in the right-most part of Fig. 1(a) \cite{Hoffmanarxiv2013}. 

The lowest level of complexity at which REB$_6$ ARPES data could be treated would then be in terms of elliptical contours due to the cross-section of the bulk ellipsoids cut at fixed $k_{\textmd{z}}$ (determined by the photon energy), and to associate the observed change in contour size to a rigid shift of the bulk band structure as illustrated in the cartoon of Fig.1(b).
If this was the whole story, then the business of ARPES investigations of the hexaborides would be over and done with after one or two beam times at a good synchrotron source. 
What gives rise to the lively scientific debate for SmB$_{6}$ and YbB$_{6}$ are two additional and interesting factors.
One is the observed surface confinement of the electron pockets at X (or rather $\overline{\textmd{X}}$), and the other has already been mentioned and is the combination of the theory prediction and the smoking guns in transport and related experiments that SmB$_6$ could harbor topological surface states due to it being a topological Kondo insulator.\\
\indent
Numerous ARPES studies agree that the electron pockets forming the elliptical contours on the Fermi surface of SmB$_{6}$(001) and YbB$_{6}$(001) do not show the strong out of plane dispersion one would expect for a bulk state of a cubic material \cite{MingSmB6,Denlingerarxiv2013,FengYbB6,MingYbB6,EmmanouilYbB6,HasanYbB6}.
For the SmB$_6$ case, as a lack of bulk, transport-active electronic states at E$_F$ are expected from the topological Kondo insulator theory and from the transport data, it is not surprising that many studies have assigned the quasi-2D states seen in ARPES of the (001) cleavage surface to new surface states which are unrelated to the bulk states and of topological character \cite{MingSmB6,FengSmB6,HasanSmB6}. Based only on the theory prediction, the surface-related states seen in some ARPES studies of YbB$_6$ have also been interpreted as being topological in nature \cite{MingYbB6,FengYbB6,HasanYbB6}.\\
\indent 
Nevertheless, similarities between these quasi-2D states at the surface and the bulk band structure cannot be put aside lightly. Consequently, a second set of ARPES studies have argued that surface-related states may occur in these systems, possibly due to mechanisms other than topological insulator behavior, for example as a result of confinement of the bulk states in a potential generated by strong surface band bending \cite{EmmanouilSmB6,EmmanouilYbB6,EmileSmB6} or the polarity of the cleavage surface \cite{DamascelliSmB6}.\\
\indent
Band bending is practically inescapable at surfaces of three-dimensional insulating or semiconducting compounds which lack a low-energy, Van der Waals-bonded cleavage plane. From band bending, 2D (surface confined) states can arise, and they can posses complex spin-momentum structures in the case that Rashba splitting becomes operative \cite{EmmPRB2010,Yaji2010,EmmJesp2010,Syro2014}.   
Arguments from transport measurements were made against accumulation-layer-related 2D transport in SmB$_6$ \cite{Wolgast2013}, and were based on single band analysis of Hall data in comparison with data from semiconductor heterostructure devices \cite{footnote1}.
Are these arguments a problem for considering band bending when interpreting ARPES data of the hexaborides?
The answer is negative, for three reasons.
Firstly, the total carrier concentration derived from de Haas van Alphen experiments on SmB$_6$ \cite{Li2013} is only 1$\times$10$^{14}$/cm$^2$. This is the same ballpark as band-bent systems can generate, which was not the case for the accumulation layer carrier numbers exceeding 10$^{15}$/cm$^2$, which would have been required to match the simpler, single-band Hall analyses of the low-T transport data \cite{Wolgast2013}. 
Secondly, the cleavage surface of a hexaboride is not a well-controlled heterostructure.
Even for simple, covalently bonded semiconductors like silicon, the fracture of a covalently-bonded network yields surface states with carrier concentrations in excess of 10$^{13}$/cm$^2$  \cite{Hasegawa2000}, well above values for silicon device structures.
Cleaving a REB$_6$ crystal necessarily involves rupture of all of the B-B covalent bonds pointing along the $z$-direction, which have an area density of order 5$\times$10$^{14}$/cm$^2$ for the (001) surface.
Thirdly, and finally, the very high surface sensitivity of ARPES does mean that if it is present, band bending will dominate the energetics (and impact the dimensionality) of any and all electronic states observable using this technique.\\
\indent
The band-bending potential offers a simple manner to combine the observed lack of out-of-plane dispersion with the in-plane electronic fingerprint very like that of a bulk 5\textit{d} state. Thus, under application of Occam's razor, the conventional surface physics driving band bending which can lead to surface-confined and even spin-polarized surface states needs to be explored and adequately excluded in the hexaborides before ARPES data can be used to argue convincingly for the topological nature of the quasi-2D states seen for these systems. \\
\begin{figure*}
\centering
\includegraphics[width =18 cm]{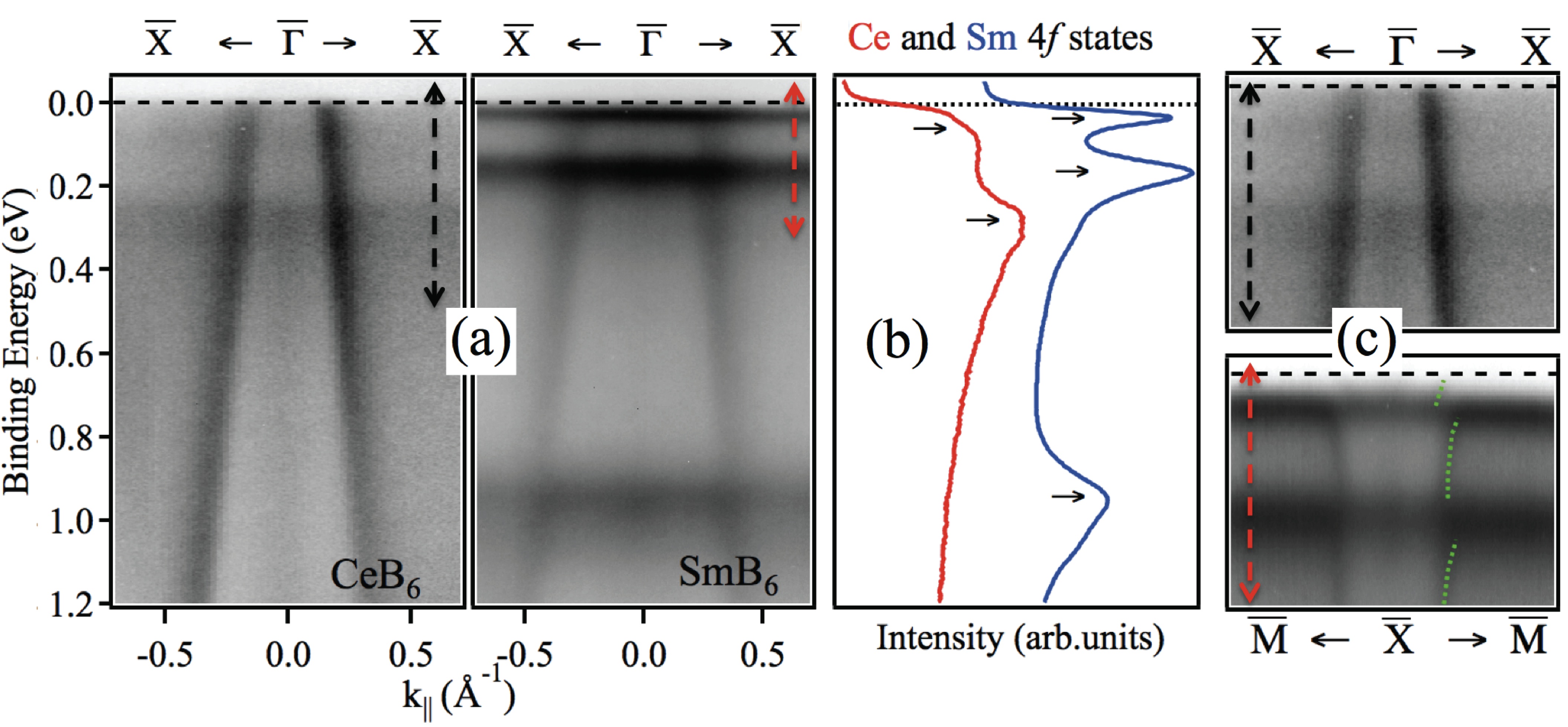}
\caption{(a) Experimental electronic structure of CeB$_{6}$(001) and SmB$_{6}$(001) along the $\overline{\Gamma\textmd{X}}$ high-symmetry direction of the surface Brillouin zone. (b) Energy distribution curves obtained by integrating the data shown in the (a) panels along the momentum direction. The small black arrows in panel (b) show the energy position of the 4\textit{f}-derived multiplets. These intense peaks are better defined in the case of SmB$_{6}$ due to the higher degree of coherence of the \textit{f} bands. (c) Near-$E_{\textmd{F}}$ dispersion of the electronic structure for CeB$_{6}$(001) (top) and SmB$_{6}$(001) (bottom) in binding energy windows shown by the double-headed arrows in panel (a).
Changes in the dispersion of the \textit{d} states due to \textit{d}-\textit{f} hybridization are observed only in the case of SmB$_{6}$ (bottom), where we show data along the $\overline{\textmd{XM}}$ high-symmetry direction. The green dotted lines illustrate the changes in the dispersion relation of the 5\textit{d} states when they cross the flat, 4\textit{f} bands. All data were acquired at 38K using a photon energy of 70 eV.}
\label{edm}
\end{figure*}
\indent    
The band structure calculations mentioned earlier \cite{Gmitra2014,FengLu2013} do not take band bending into account, and this could be a natural explanation of the different chemical potential in the calculations (i.e. $E_{\textmd{F}}$ in a bulk energy gap), compared with experiment (i.e. $E_{\textmd{F}}$ crossing states which are strongly reminiscent of the states in the bulk).
We note that early photoemission studies on metal hexaborides \cite{DenlingerPRL2002,MorimotoPRB} did indeed observe different energy values for the surface and bulk chemical potential. We close our discussion on the dispersive states by noting that the strong similarities between the three different compounds do argue for a generic interpretation for the electronic structure of the outermost nanometer of cleaved crystals of REB$_{6}$ materials, such as that provided by the band bending picture. 

Fig. 2 allows further comparison of the near-$E_{\textmd{F}}$ electronic structure of RE hexaborides. Panel (a) shows the experimental energy dispersion of CeB$_{6}$ and SmB$_{6}$ along the $\overline{\Gamma\textmd{X}}$ high symmetry direction of the SBZ. The temperature (38K) and all other experimental conditions were identical for both datasets. In both panels, one can readily see dispersive and essentially non-dispersive states.
The dispersive states have been discussed above, and are the RE 5\textit{d} bands which yield the elliptical contours at $E_{\textmd{F}}$.
The flat bands are of 4\textit{f} origin, indicated schematically with light blue lines in Fig.1(b). 
In SmB$_{6}$ emission from three \textit{f} states are observed at binding energies of 40, 170 and 960 meV. These states are the 6H$^{5/2}$, 6H$^{7/2}$ and 6F final state multiplets of the 4\textit{f}$^{6}$ $\rightarrow$ 4\textit{f}$^{5}$ transition
 \cite{SoumaPhysicaB2002,Neupanearxiv2013,DenlingerPhysicaB}. In CeB$_{6}$, the observed \textit{f} states lie grouped at 280 meV and 50 meV binding energy, and they have been attributed to the 4\textit{f}$^{1}$ final state \cite{Hasanarxiv2014,SwapnilJPCM,SwapnilAPL} . 

Comparing the two sets of \textit{f}-derived features, one can readily see significant differences in their relative amplitude and energy width.
In SmB$_{6}$ the intensity of the \textit{f} multiplets dominates the spectrum, with the energy width of the multiplets decreasing with the binding energy [see the blue energy distribution curve in Fig.2(b)]. On the other hand, the \textit{f} states in CeB$_{6}$ are rather broad [see the red energy distribution curve in Fig.2(b)], and of much lower intensity in comparison to the \textit{d}-derived features. One possibility is to attribute differences in the relative spectral weight of the $f$ states to their different occupancy, and the lower  Nevertheless, we also note that the $f$-state spectral weight in SmB$_{6}$ has been previously shown to be strongly temperature-dependent \cite{Denlingerarxiv2013}. As a matter of fact, energy broadening and reduced spectral intensity, as observed in CeB$_{6}$, are reminiscent of the RE 4\textit{f} emission in SmB$_{6}$ at high temperature \cite{Denlingerarxiv2013}, in which combined ARPES and DMFT studies have shown that the incoherent 4\textit{f} states do not hybridize with the RE 5\textit{d} bands \cite{Denlingerarxiv2013}.
This is in stark contrast to the low-temperature regime, in which well-defined $f$ bands induce clear hybridization effects in the dispersion relation of the RE 5\textit{d} states \cite{Denlingerarxiv2013,EmmanouilSmB6}.
In CeB$_{6}$, the lack of significant hybridization-related spectroscopic fingerprints at all temperatures make this system a good analogy to the high-temperature regime of SmB$_{6}$. As an example of this, the upper panel of Fig. 2(c) shows that the weak and energy-broad RE 4\textit{f}  states in CeB$_{6}$ do not affect the energy dispersion of the RE 5\textit{d} states. The latter penetrate through the 4\textit{f} emission, while remaining unperturbed up to $E_{\textmd{F}}$. In other words, seen from the point of view of ARPES data, there are no \textit{d}-\textit{f} hybridization effects in CeB$_{6}$ at 38K.
This is in contrast to SmB$_{6}$ where the bottom panel of Fig. 2(c) shows clear hybridization effects observed at the same temperature. The hybridization is strongest for the $d$-states interacting with the \textit{f} state that lies closest to the $E_{\textmd{F}}$. This observation agrees with the fact that the \textit{f}-derived feature at lowest binding energy has the smallest energy width [see again the blue energy distribution curve in Fig. 2(b)]. All spectroscopic fingerprints and indications of \textit{d}-\textit{f} hybridization (small bandwidth and high intensity of the \textit{f} states, $k$ discontinuities, changes of slope) are absent in the case of CeB$_{6}$, which is in line with its valence of three, and 4$f$ occupation of unity.  

The lack of strong \textit{d}-\textit{f} interaction in CeB$_{6}$ suggests that the bulk Ce 5\textit{d} states are unchanged all the way up to $E_{\textmd{F}}$ [Fig. 2(c)]. Indeed Fig. 1(e) shows that the in-plane Fermi surface contours of CeB$_{6}$(001) are reminiscent of the bulk 5\textit{d} ellipsoids sketched in the rightmost panel of Fig. 1(a). Knowing that ARPES data suggest surface-confinement of the near-$E_{\textmd{F}}$ electronic states in SmB$_{6}$ \cite{MingSmB6,FengSmB6,Denlingerarxiv2013} and YbB$_{6}$ \cite{EmmanouilYbB6,MingYbB6,HasanYbB6}, the question arises whether the analogous states in CeB$_{6}$ are 3D- or 2D-like.
We therefore close the comparison of the electronic dispersion in these three RE hexaborides with ARPES data on the dimensionality of the Fermi surface contours in CeB$_{6}$(001). ARPES measurements with different photon energies probe the energy \textit{vs.} the out-of-plane component of the momentum (i.e. $k_{\textmd{z}}$), and can thus reveal the dimensionality of the electronic states. Surface-localized states will show very little or no variation on changing $h\nu$, while bulk bands are expected to disperse along all dimensions of $k$-space forming closed ($k_{\textmd{z}}$, $k_{\textmd{x}}$) and ($k_{\textmd{z}}$, $k_{\textmd{y}}$) contours.

\begin{figure}
\centering
\includegraphics[width =9.3 cm]{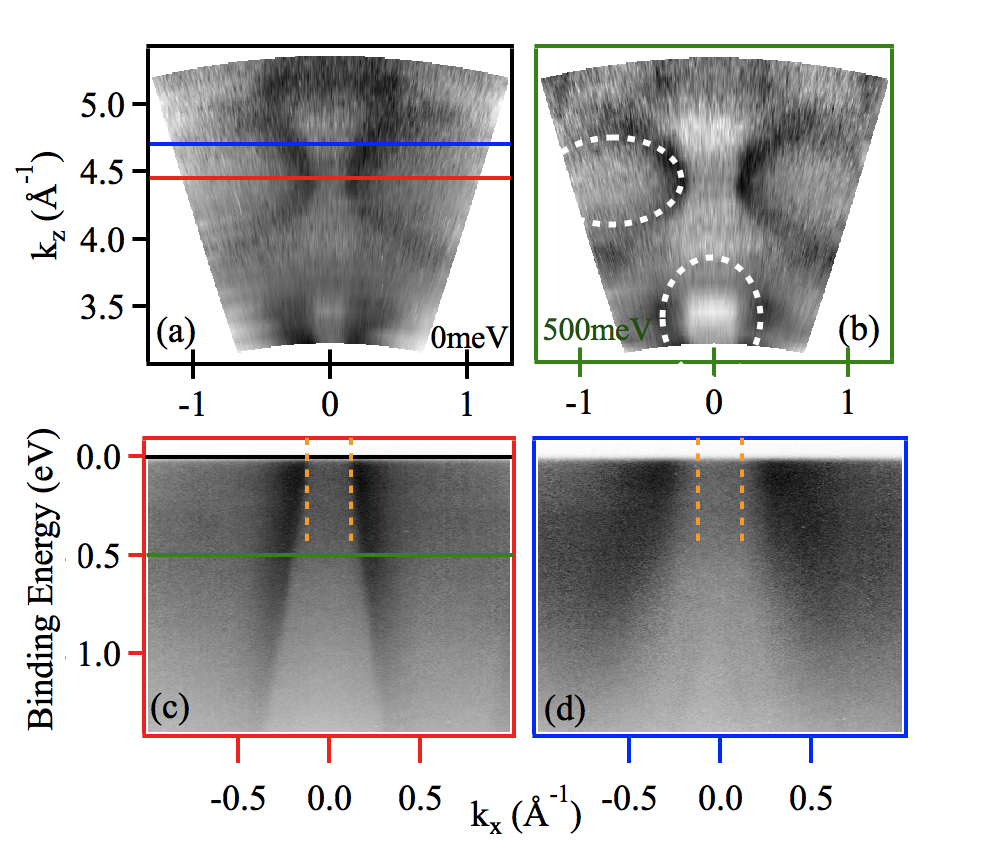}
\caption{(a), (b) Experimental momentum distributions for the occupied electronic states of CeB$_{6}$(001) in the $k_{\textmd{z}}$-$k_{\textmd{x}}$ plane. $k_{\textmd{x}}$ is along $\overline{\Gamma\textmd{X}}$ and $k_{\textmd{z}}$ is out-of-plane. Data is shown for two different electron energies: (a) at $E_{\textmd{F}}$ and (b) at 500 meV below the Fermi level.
These energies are indicated using black and green lines in panel (c).
The white dashed ellipses in (b) are guides to the eye that follow the experimental contours and reveal the three-dimensional character of the states. Panels (c), (d) show the dispersion of the occupied states for CeB$_{6}$(001) along $\overline{\Gamma\textmd{X}}$ at two different $k_{\textmd{z}}$ values, the latter indicated in panel (a) using the blue and red lines. Data in this figure were acquired at 38K by varying $h\nu$ between 30 and 100 eV.}
\label{kz}
\end{figure}

\begin{figure}
\centering
\includegraphics[width =8.5 cm]{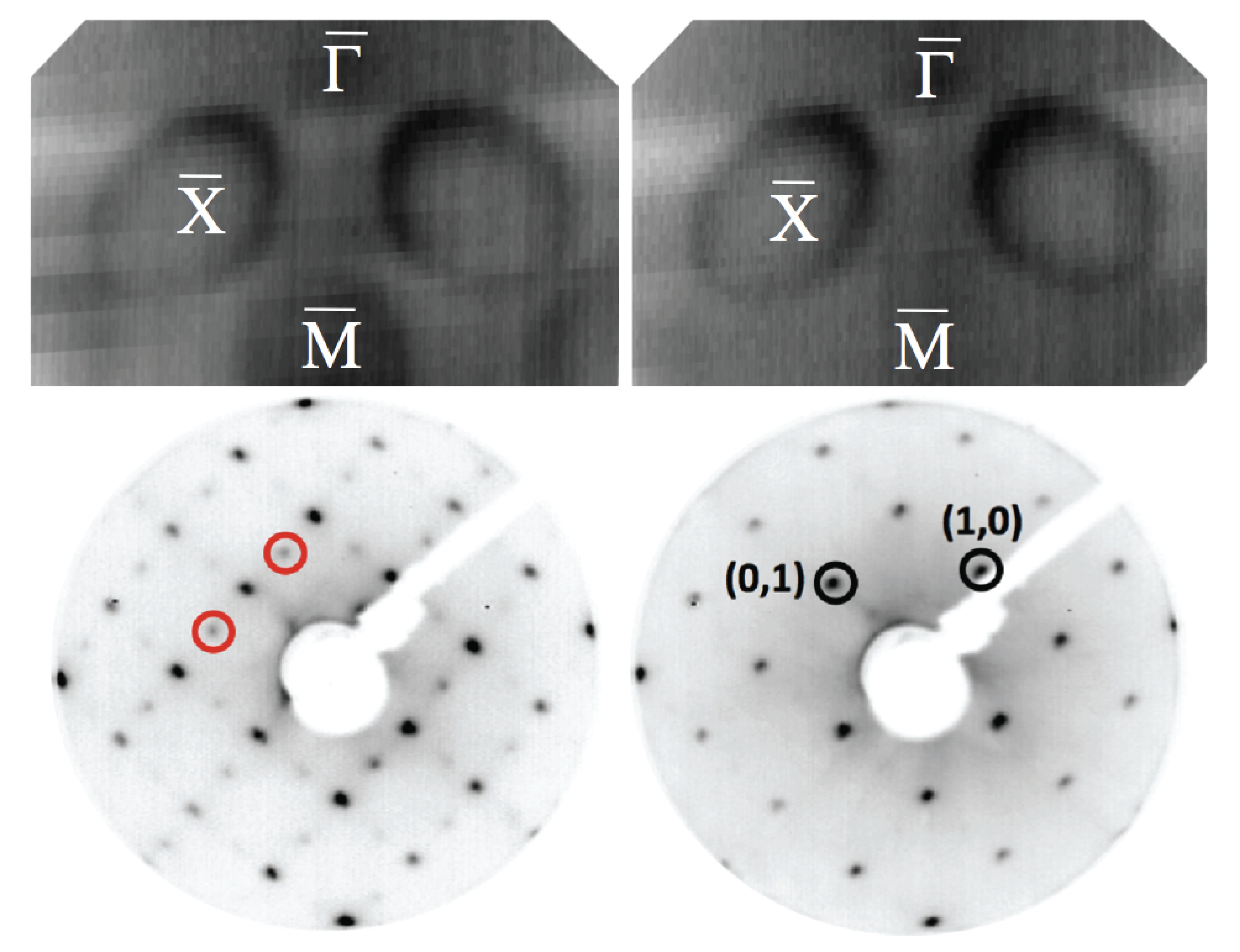}
\caption{Left images: LEED and ARPES data from freshly UHV-cleaved SmB$_{6}$(001). The LEED image shows $2\times1$ superstructure spots (highlighted with red circles) and the ARPES data contains back-folded contours at the $\overline{\textmd{M}}$ point. These two measurements reveal long-range order with $2\times1$ and $1\times2$ periodicity at early stages after cleavage. Right images: LEED and ARPES data acquired at the same sample locations five hours after cleavage, showing no traces of long-range $2\times1$ order, but rather only $1\times1$ bulk-like periodicity. ARPES results were acquired using a photon energy of 70 eV. Both LEED and ARPES data were acquired at 38K.}
\label{stm}
\end{figure}

\begin{figure*}
\centering
\includegraphics[width =17.5 cm]{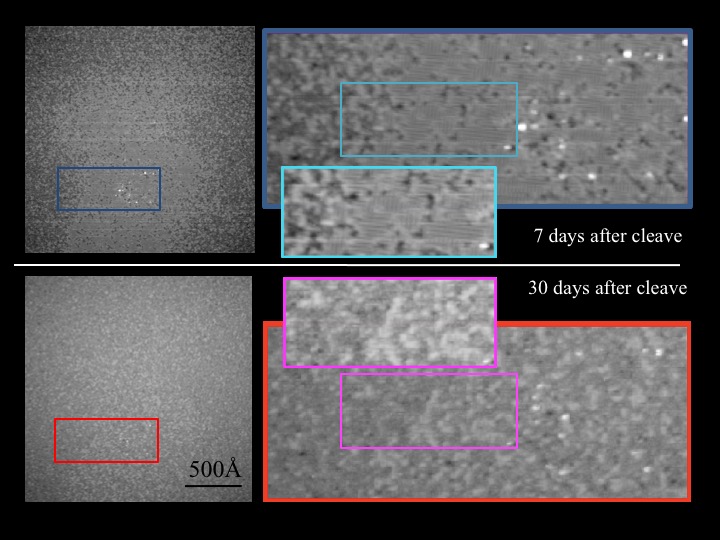}
\caption{Top row of images: STM topographs acquired 7 days after cleavage show that by this time the $2\times1$ reconstructed regions at the surface are only short-ranged: the outermost layer of the sample surface is mostly disordered. Bottom row: there are no traces of the $2\times1$ reconstruction (either long- or short-range) at any locations sampled when the STM measurements are repeated 30 days after initial cleavage. At all times the sample remained in UHV ($P$ lower than 5.0$\times$10$^{-10}$ mbar) at a temperature under 20K. All STM data were acquired at 4K.}
\label{stm}
\end{figure*}

Figs. 3(a) and 3(b) show the ($k_{\textmd{z}}$, $k_{\textmd{x}}$) contours for CeB$_{6}$(001) at two different binding energies, namely at the Fermi energy for panel (a), while panel (b) is obtained 500 meV below $E_{\textmd{F}}$.
$k_{\textmd{x}}$ is oriented along the $\overline{\Gamma\textmd{X}}$, the direction in the SBZ probed in the CeB$_{6}$ data of Figs. 2a and 2c.

The dashed ellipses in Fig. 3(b) highlight the observed experimental constant energy contours, and they can be seen to be very similar to the in-plane $I$($k_{\textmd{x}}$,$k_{\textmd{y}}$) contours shown in Fig. 1(e) (marked in that figure with a red dahed line). Putting together Figs.1(e), 3(a) and 3(b), one can readily conclude that the observed electronic states of CeB$_6$ disperse along all directions of $k$-space and that their contours at all energies in any 2D k-plane chosen from $k_{\textmd{x}}$,$k_{\textmd{y}}$,$k_{\textmd{z}}$ are ellipses.
This means that the Fermi surface in CeB$_{6}$(001) is formed by states which are not confined to the (near-)surface region of the crystal, quite unlike the situation in SmB$_{6}$(001) and YbB$_{6}$(001). Panels (c) and (d) of Fig. 3 underline the three-dimensionality of the electronic states also at the surface of CeB$_{6}$ in that they show the markedly different energy dispersion of the Ce 5\textit{d} states at two different $k_{\textmd{z}}$ values [indicated by the colored horizontal lines in panel (b)].
To guide the eye, the orange dashed lines mark the approximate $k_{\textmd{F}}$-values of the Ce 5\textit{d} states in panel (c) and these are reproduced in panel (d) for comparison.
The lack of surface confinement in the electronic band structure of CeB$_{6}$ is wholly in line with the absence of surface band bending, with the result that the electronic structure at the surface and in the bulk are identical, which was first discussed in the context of Fig. 1(b).

The comparison of the electronic structure between the three RE hexaborides made on the basis of ARPES data presented in Figs. 1-3 reveals interesting similarities (orbital character and contour shape near the Fermi level) but also crucial differences (energy position of the chemical potential \textit{vs.} states of \textit{d} and \textit{f} character, coherence of \textit{f}-derived bands, dimensionality of states at $E_{\textmd{F}}$). 
In the remaining part of this study, we will focus on SmB$_{6}$. On the top of its widely studied properties due to its fluctuating 4f valence, in the following we will reveal an interesting aspect of its complex surface structure through the combination of data from ARPES, LEED and STM.

SmB$_{6}$, like the other RE hexaborides does not possess a natural cleavage plane due to its CsCl crystal structure [Fig. 1(a)]. This means that when the sample is fractured to reveal a (001) plane, simple (but nonetheless persuasive) electrostatics would lead to cleavage surfaces that are - on average - an equal combination of Sm- and B$_{6}$-termination. Complete Sm$^{2+}$(001) or B$_{6}$$^{2-}$(001) terminations would be polar and highly energetically unfavorable \cite{DamascelliSmB6,Hoffmanarxiv2013}, unless they are stabilized by means of an electronic or structural reconstruction. 
If every second Sm ion were to remain on either side of the cleave, then this could give a $\sqrt{2}$$\times$$\sqrt{2}$ or a combination of 2$\times$1 and 1$\times$2 reconstructions.
A 2$\times$1 surface reconstruction has been indeed observed on SmB$_{6}$(001). This reconstruction has been found on UHV (ultra high vacuum) annealed samples by LEED \cite{MiyazakiSmB6,AonoSmB6} and on UHV cleaved samples by STM topographic imaging \cite{Hoffmanarxiv2013,Rossler2014} and also in one ARPES study \cite{MingSmB6}.
However, for UHV cleaved samples, neither LEED \cite{EmmanouilSmB6,DamascelliSmB6} nor the vast majority of ARPES studies \cite{FengSmB6,EmmanouilSmB6, Denlingerarxiv2013,DamascelliSmB6,HasanSmB6} have found signs of a 2$\times$1 reconstruction.
In addition, STM studies have shown that that 2$\times$1 (and there equivalent 1$\times$2) is not the only periodicity observed on the surface of UHV cleaved SmB$_{6}$(001) \cite{Hoffmanarxiv2013,Rossler2014,Ruan2014}. It is therefore clear that there is an open debate on the true surface structure of cleaved SmB$_{6}$(001).

Here, we provide new insight into the surface periodicity of UHV cleaved SmB$_{6}$(001) by reporting - in Figs. 4 and 5 - time dependent changes in the surface symmetry under ultra high vacuum conditions. Both our LEED and ARPES data on freshly cleaved surfaces [Fig. 4, left-hand images] show clear signs of long-range 2$\times$1 periodicity at some regions of the sample surface. This is evident in the surface diffraction data from the additional spots in the LEED pattern that would have fractional indices based on the 2D reciprocal lattice of the 1$\times$1 structure. In the ARPES data the larger real space unit cell at the surface results in the folding of bands at the new (smaller) surface Billouin zone boundary, and this leads to clear ARPES intensity at the ($\pi$/$\pi$) location in reciprocal space ($\overline{\textmd{M}}$), which is expected to be free of bands for both the bulk and the 1$\times$1 surface structure. This extra feature in the left-hand Fermi surface map in Fig. 4 at $\overline{\textmd{M}}$ is, then, a diffraction replica of states at the $\textmd{X}$ points, due to the 2$\times$1 and 1$\times$2 surface superstructure. 

In the same LEED and ARPES experiments, other parts of the fresh cleavage surfaces showed only 1$\times$1, bulk-like periodicity. Interestingly, 5 hours after the cleave, the regions that originally exhibited long-range 2$\times$1 periodicity, no longer showed the 2$\times$1 superstructure spots in LEED, and nor did they exhibit 2$\times$1 back-folding of the bands in ARPES. The right-hand LEED and Fermi surface images shown in Fig. 4, show no features that cannot be simply explained by the 1$\times$1 structure. This means that on the timescale of 5 hours after cleavage, no sign of a long-range $2\times1$ periodicity could be found by either of these experimental techniques and all sample surface locations probed showed an apparent $1\times1$ bulk-like termination. We note that in the case of changes induced by residual gas atoms, the 5-hour timescale will depend on the base pressure of the system, which in our case was lower than 5.0$\times$10$^-10$ mbar.
The LEED and ARPES data presented here average over length scales of $\sim$1 mm and 100 $\mu$m. Thus, it could be that at longer times after 
cleavage, the $2\times1$ period superstructure is still present, but is primarily short-range ordered.

To examine this possibility, we turn to STM as an effective local probe of the surface structure.
The upper row of images in Fig. 5 shows typical STM topographic data for UHV-cleaved SmB$_{6}$(001) acquired 7 days after UHV cleavage. The surface is clearly a mixture of disordered regions and regions with short-range $2\times1$ and $1\times2$ periodicity, as highlighted by the progressive zooms (left-hand square image then dark blue oblong then light blue oblong). 
The light-blue zoomed region (of dimension 372$\times$154 \AA$^2$) 
shows a transition between a short-range ordered superstructure region on the right and a more disordered/amorphous structure on the left, both of which are more visible in the brightness/contrast-tuned inset of the light blue zoomed area. 
The topographs in the upper row of Fig. 5 suggest, therefore, that after several days in UHV, the $2\times1$ (and $1\times2$) regions at the surface are only of order 100 nm in lateral size and thus are too small to yield the visible superstructure spots in LEED or well-defined umklapp features in ARPES shown in the left-hand data images in Fig. 4.
We note that 1$\times$1 LEED images are still recordable from UHV-cleaved SmB$_6$ after UHV exposure on the timescale of a small number of days.   
The question is, then, how should we look at the time dependence reported in Figs. 4 and 5?
The longer-range superstructure order seen for portions of the fresh cleave, becomes shorter-ranged - and thus invisible to LEED but still detectable using STM - on the time scale of several days (in UHV and low temperature).   
Mass transport away from the sample or long-range movement of atoms across the surface would seem an unlikely scenario to explain these observations, given the temperature of the sample never exceeded 20K.
A more likely scenario could involve the relaxation of the surface superstructure, possibly mediated by adsorption of residual gases from the vacuum.      
As a test of this hypothesis, STM topographs were re-recorded after a further 30 days in UHV (with sample temperatures held below 20K, and without changing location).
The lower row of images shown in Fig. 5 show the topographic data. 
Careful comparison of the images in the upper and lower image rows of Fig. 5 yields that there are sufficient (probably defect-related) features in the topographs (showing up as white or black dots/blobs), to enable re-location of the same region of the cleaved surface, also after 30 days.
The data clearly show continued relaxation of the $2\times1$ and $1\times2$ features on the cleavage surface, so as to give a long-term cleavage surface that - from the point of view of the exquisite surface sensitivity of STM - appears to be without crystalline structure; looking similar to the non-ordered parts of the surface imaged after 7 days in the upper set of topographic data. 
30 days after cleavage, also on shifting the STM's field of view to wholly different regions of the cleave did not uncover topographic contrast any more crystalline than that seen in the lower images of Fig. 5. 
These STM data could be seen to beg the question whether the clear and strong $1\times1$ periodicity seen by LEED and ARPES up to a few days after the cleave (and by extension also possibly the $1\times1$ signal often seen in ARPES directly after cleavage) could, in fact, originate from the first unfractured layer underneath the crystal termination, rather than from the strongly disordered very outermost layer of the crystal imaged using the STM as shown in Fig. 5. \\
\indent
Taken together, the data presented in Figs. 4 and 5 make it clear that the details of the surface structure of cleaved SmB$_6$ are far from simple and far from static, even in UHV. 
On the one hand, these data help explain the differences between data from highly local and more spatially averaging techniques on the surface structure of SmB$_{6}$ by underlying the importance of the time interval between cleavage and measurement, as well as the spatial sensitivity of the probe.
On the other hand, they also provide clear motivation for more quantitative analysis of the (3D) surface structure of real cleaves of SmB$_6$, as it is ARPES data from these kinds of surfaces that are being compared to the perfect terminations in slab-based calculations.    
On the theory-front, first steps are being taken to address possible deviation of the surface from the bulk. Specifically, changes in the energetics and dispersion relation (effective mass) of the predicted Dirac cone for SmB$_6$ have been calculated due to so-called Kondo breakdown at the surface of the material \cite{Alexandrov2015K}.\\

\section{CONCLUSIONS}

We performed a comparative ARPES study of the electronic structure of three RE hexaborides: YbB$_{6}$(001), SmB$_{6}$(001) and CeB$_{6}$(001). We discuss similarities and differences observed in the spectroscopic fingerprints of the RE-derived 5$\textit{d}$ and 4$\textit{f}$ electronic states among the different compounds. We argue that the main features of the experimental band structure reflect changes in the RE valence in the form of  an energy shift of the RE 5$\textit{d}$ electronic states with respect to the Fermi level. The occupancy of the $\textit{d}$ states is highest for CeB$_{6}$, in which the RE is trivalent and lowest for YbB$_{6}$, in which the RE is divalent. The mixed valent case of SmB$_{6}$ falls nicely in between. As a result, the Fermi surface contours observed on the (001) surfaces of these three compounds vary in size, but not in their overall shape. 
In contrast, the dimensionality of the electronic states at $E_{\textmd{F}}$ differs within these three members of the hexaboride family.
The electronic states observed for YbB$_{6}$(001) and SmB$_{6}$(001) are surface confined. In the former, a straightforward band-bending scenario is able to explain the data admirably, in keeping with the simple divalent character of Yb in this system.
For SmB$_6$, band bending at the surface is also an option to be taken seriously. However, the mixed valence of the Sm, the strongly temperature dependent behavior of the 4\textit{f}-5\textit{d} hybridisation and the very low energy scales of the energy gaps expected from hybridisation still makes this system very challenging, even for modern high-resolution ARPES experiments.   
From the point of view of its low-lying electronic states seen by ARPES, CeB$_{6}$(001) is as straightforward as its Yb counterpart. We observe a clear three-dimensional character of the Fermi surface contours for CeB$_{6}$(001), which fits the expectation that neither band bending nor surface confinement are relevant for this 3D, metallic system.
CeB$_6$ provides a nice example of the spectroscopic signatures of a system with low coherence of the RE 4$\textit{f}$ states - which gives 4\textit{f} multiplets that are broad in energy with relatively weak spectral weight - and with negligible hybridization between the \textit{d} and \textit{f} states. This results in an unperturbed dispersion of the 5$\textit{d}$ states, going as it were right `through' the 4f levels, also at low temperature. 
Both these signatures can also be found in SmB$_6$, but only at high temperatures.\\ 

All in all, these experiments show the suitability of high-resolution ARPES to provide new insight into the electronic structure and properties of strongly correlated RE hexaborides, providing information on the energy dispersion, momentum distribution, dimensionality and coherence of the electronic states. Finally, we show that when ARPES is combined with scanning tunnelling microscopy and electron diffraction techniques, additional structural information can be obtained, and in the case of cleaved SmB$_{6}$(001), this reveals a time-dependent relaxation of the surface superstructure, even under ultra high vacuum and cryogenic conditions.\\

\section{ACKNOWLEDGEMENTS}
This work is part of the research programme of the Foundation for Fundamental Research on Matter (FOM), which is
part of the Netherlands Organization for Scientific Research (NWO).
The research leading to these results has also received funding from the European Community's Seventh Framework Programme
(FP7$/$2007-2013) under grant agreement n$^{\textmd{o}}$ 312284 (CALIPSO).
We are grateful to X. Zhang and coworkers
for the provision of single crystals in the initial stage of this study and to A. Varykhalov for experimental support during the experiments at HZB-BESSY.


\begin{thebibliography}{70}
\bibitem[{foo()}]{footnote1}
\bibinfo{note}{There are no such statements in the literature for YbB$_6$ as
  this system is not insulating and shows no anomalous transport behavior.}
\end{thebibliography}
\end{document}